# PROPAGATION OF PHOTON IN RESTING AND MOVING MEDIUM

## J. Zaleśny

Institute of Physics, Technical University of Szczecin,

Al. Piastów 48, 70-310 Szczecin, Poland

**A b s t r a c t**

The propagation of photon in a dielectric may be described with the help of the scalar and vector potentials of the medium. The main novelty of the paper is that the concept of the vector potential (which is connected with the velocity of the medium) can be extended to relativistic velocities of the medium. The position-dependent photon wave function was used to describe the propagation of the photon. The new concepts of the velocity of photon as particle and the photon mass in the dielectric medium were proposed.

PACS numbers: 03.50.De, 03.65.w

# 1. INTRODUCTION

Consider a photon in a dielectric medium. But what photon really is? In modern physics photon is nothing more than quantum excitation of the electromagnetic field. We have learned from quantum electrodynamics, that in the dielectric, in fact, there are no photons but polaritons, i.e. excitations of electromagnetic field coupled to the medium. However a different point of view is also possible. Remember the case of electron in an external electromagnetic field. On the first quantization level the electron is treated as a quantum object moving in the classical field. In this paper I develop a similar description for a photon. The photon treated as a quantum object 'feels' the medium as an external classical field. To describe a photon in terms of a one-particle wave function, i.e., on the *first* quantization level, I follow the way presented in [1-3]. Another approach was proposed in [4]. The concept of the position-representation photon wave function has a long history and is still controversial. Nevertheless I do not want to discuss the question here. The reader interested in this problem is referred to [1–4] and references therein.

The paper is organized in the following way. In Section 2, I develop the description of a photon in medium in terms of the photon wave function. Some attempts of this kind have been presented in [1-3]. What is new in my approach is to show that the influence of the medium on the photon can be described through some potentials. Though generally the idea is not new, see e.g. [5-6], but here I realize it within the formalism of the photon wave function. On this basis, in Section 3, the nonzero mass of photon and the concept of the velocity of photon as a particle appear in a quite natural way. The velocity of photon is different from the phase or group velocity and, up to my knowledge, is a new concept.

I show in Section 4 that the motion of the dielectric can be connected with the optical analog of the vector potential. This idea has been already presented in literature, see papers [7-8]. What is new here is that the concept of the vector potential of the medium can be extended for relativistic velocities of the medium. With the help of the scalar and vector potentials of the medium one can define some optical analogs of electric and magnetic fields, and the optical analog of the Lorentz force (acting on the photon in the medium). The potentials are gauge fields and the analogs of electric and magnetic fields are gauge invariant.

## 2. THE SCALAR POTENTIAL OF MEDIUM

In papers [1-3] the following form of the Schrödinger equation for free photon was proposed

$$i\hbar \partial_t F = H_f F .\qquad(1)$$

$$F = \begin{bmatrix} \mathbf{E}(t,\mathbf{r}) + i\,\mathbf{H}(t,\mathbf{r}) \\ \mathbf{E}(t,\mathbf{r}) - i\,\mathbf{H}(t,\mathbf{r}) \end{bmatrix} \quad;\quad H_f = c\begin{bmatrix} \mathbf{p}\cdot\mathbf{S}, & 0 \\ 0, & -\mathbf{p}\cdot\mathbf{S} \end{bmatrix} \quad;\qquad(2)$$

$\mathbf{p} = -i\hbar\nabla$ - momentum of photon;

$(S_i)_{kl} = -i\varepsilon_{ikl}$ - spin photon matrix ($\varepsilon_{ikl}$ - antisymmetric Levi-Civita symbol ).

On the classical language, the equations are equivalent to the following Maxwell equations

$$\partial_t \mathbf{E} = c\,\nabla\times\mathbf{H} \;;\; \partial_t \mathbf{H} = -c\,\nabla\times\mathbf{E} \;;\; \mathbf{D} = \mathbf{E} \;;\; \mathbf{B} = \mathbf{H} ,\qquad(3)$$

describing free fields in vacuum. Since all the information carried by function $F$ is contained in its positive energy (positive frequency) part $F^{(+)}$, following [3], I take this part as the true photon wave function and denote it as $\psi$

$$\psi = F^{(+)}\qquad(4)$$

To become a complete set of Maxwell equations, eq.(3) must be supplied by divergence conditions $\nabla\cdot\mathbf{E} = 0$, $\nabla\cdot\mathbf{H} = 0$. It is equivalent to the relation $\mathbf{p}\cdot\psi = 0$.

In order to describe the propagation of photon in dielectric, one should include in Hamiltonian the interaction term. On the microscopic level, such interaction is rather complicated, but here I will take it into account in a phenomenological way. Let us begin with stationary states of the photon in a homogeneous dielectric. For a stationary state the wave function takes on the form

$$\psi_\omega = \varphi_\omega(\mathbf{r})\exp(-i\omega t) , \quad \text{where} \quad \varphi_\omega(\mathbf{r}) = \begin{bmatrix} \mathbf{E}(\mathbf{r}) + i\mathbf{H}(\mathbf{r}) \\ \mathbf{E}(\mathbf{r}) - i\mathbf{H}(\mathbf{r}) \end{bmatrix} \tag{5}$$

The propagating photon in every time and in every space point, 'feels' the same coupling with the medium. We may try to describe the interaction by a single constant coupling value $\Omega_\omega$. If some inhomogeneities are in the dielectric, then the interaction depends on **r** and will be modeled by a function $\Omega_\omega(\mathbf{r})$. Sometimes, I want to restrict considerations to the non-magnetic media, i.e., $\mu = 1$. It means that the medium is coupled only to the electric part of the photon wave function. Generally, the couplings of the medium with the electric and magnetic part of the wave function may be different. To take it into account I introduce two real and symmetric matrices $\gamma$ and $\eta$ which split the wave function $\psi_\omega$ into electric and magnetic parts

$$\gamma \psi_\omega + \eta \psi_\omega = \psi_\omega , \tag{6}$$

$$\gamma \psi_\omega = \begin{bmatrix} \mathbf{E} \\ \mathbf{E} \end{bmatrix} \quad \text{and} \quad \eta \psi_\omega = \begin{bmatrix} i\mathbf{H} \\ -i\mathbf{H} \end{bmatrix} , \tag{7}$$

$$\eta = \frac{1}{2}\begin{bmatrix} 1 & -1 \\ -1 & 1 \end{bmatrix} \quad \text{and} \quad \gamma = \frac{1}{2}\begin{bmatrix} 1 & 1 \\ 1 & 1 \end{bmatrix}. \tag{8}$$

The projection operators $\gamma, \eta$ fulfill the following relations i.e.: $\gamma^2 = \gamma$, $\eta^2 = \eta$, $\gamma\eta = 0$. Thus the case of the propagating photon in inhomogeneous nonmagnetic dielectric can be modeled in the following way

$$\hbar\omega \psi_\omega = H_f \psi_\omega - \Omega_\omega(\mathbf{r}) \gamma \psi_\omega . \tag{9}$$

I interpret the term $\Omega_\omega(\mathbf{r}) \gamma$ as a *potential energy* operator of the photon in dielectric. In order to see what is the meaning of the quantity $\Omega_\omega(\mathbf{r})$ in the classical language, one may translate equation (9) into the ordinary form of non-vacuum Maxwell equations

$$-i\omega[1 + 4\pi\chi_\omega(\mathbf{r})]\mathbf{E} = c\nabla\times\mathbf{H} \quad ; \quad i\omega\mathbf{H} = c\nabla\times\mathbf{E} , \qquad (10)$$

$$\text{where} \qquad \chi_\omega(\mathbf{r}) = \frac{1}{4\pi}\frac{\Omega_\omega(\mathbf{r})}{\hbar\omega} .$$

Thus $\Omega_\omega(\mathbf{r})$ is directly connected with the dielectric susceptibility $\chi_\omega(\mathbf{r})$.

It is easy to generalize this approach and write Maxwell equations for dispersive media in a form of Schrödinger equation. If simultaneously many frequencies are present in the medium, then it is reasonable to expect (from the quantum mechanical point of view - indistinguishable alternatives) that the interaction of medium with the photon in such a *non-stationary* state is described by superposition of single-frequency interaction terms. Thus, if the photon is described by a wave packet $\psi(t,\mathbf{r})$, then the interaction of the packet with the medium can be described by an integral operator $\hat{\Omega}_L$ in the following way

$$\hat{\Omega}_L \psi(t,\mathbf{r}) = \int \Omega_\omega(\mathbf{r}) \psi_\omega(t,\mathbf{r}) d\omega = \int \Omega_\omega(\mathbf{r}) \varphi_\omega(\mathbf{r}) \exp(-i\omega t) d\omega . \qquad (11)$$

The equation of motion of the photon takes on the quasi-Schrödinger form

$$i\hbar\partial_t \psi = H_f \psi - \gamma \hat{\Omega}_L \psi . \qquad (12)$$

The integral form of the interaction term makes the relation between fields **D** and **E** nonlocal in time, i.e.,

$$\mathbf{D}(t,\mathbf{r}) = \mathbf{E}(t,\mathbf{r}) + 4\pi \int \chi(\tau) \mathbf{E}(t-\tau,\mathbf{r}) d\tau . \qquad (13)$$

Equation (12) simplifies in some special cases. E.g. in a *non-dispersive* medium

$$\Omega_\omega(\mathbf{r}) = 4\pi\chi(\mathbf{r})\hbar\omega , \qquad (14)$$

where $\chi(\mathbf{r})$ does not depend on $\omega$ the equation (12) becomes

$$i\hbar\,\partial_t (1 + 4\pi\chi(\mathbf{r})\gamma)\psi = H_f \psi . \qquad (15)$$

In this case it is possible to construct some *effective* wave function (and Hamiltonian) and such effective form is used in [1-3].

Another interesting case is when $\Omega(\mathbf{r})$ is *independent* of $\omega$, then $\hat{\Omega}_L \psi(t,\mathbf{r}) = \Omega(\mathbf{r})\psi(t,\mathbf{r})$. In this case the similarity to the case of electron in an external field is the most appealing.

For simplicity, in the next Sections, I restrict discussion to stationary states.

## 3. THE MASS AND VELOCITY OF PHOTON IN THE MEDIUM

When the couplings with the electrical and magnetic parts are taken into account, the Schrödinger equation takes on the form

$$\hbar\omega\,\psi_\omega = H_f \psi_\omega - \Omega_\omega(\mathbf{r})\,\gamma\,\psi_\omega - \Gamma_\omega(\mathbf{r})\,\eta\,\psi_\omega. \tag{16}$$

$\Omega_\omega(\mathbf{r})$, $\Gamma_\omega(\mathbf{r})$ have interpretation of potential energies. $\Gamma_\omega(\mathbf{r})$ is connected with magnetic susceptibility $\chi^m_\omega(\mathbf{r}) = (1/4\pi)\,\Gamma_\omega(\mathbf{r})/\hbar\omega$. In order to obtain the connection between the *total energy* and *momentum* one may iterate this equation. In the case of a homogeneous medium one obtains

$$(\hbar\omega)^2\,\psi_\omega = \left[H_f^{\,2} - (\hbar\omega\Omega_\omega + \hbar\omega\Gamma_\omega + \Omega_\omega\Gamma_\omega)\right]\psi_\omega, \tag{17}$$

where the identities $H_f\,\gamma + \gamma\,H_f = H_f$, $H_f\,\eta + \eta\,H_f = H_f$ have been used.

The equation (17) is in fact the classical wave equation. It is easy to note that

$$\frac{\Omega_\omega}{\hbar\omega} + \frac{\Gamma_\omega}{\hbar\omega} + \frac{\Omega_\omega\Gamma_\omega}{(\hbar\omega)^2} = \varepsilon_\omega\mu_\omega - 1 = n_\omega^2 - 1, \tag{18}$$

where $\varepsilon_\omega$, $\mu_\omega$ are permittivity and permeability of the medium, and $n_\omega$ is refractive index, and also that

$$H_f^{\,2} = c^2(\mathbf{p}\cdot\mathbf{S})^2 = c^2 p^2\,;\qquad (\mathbf{p}\cdot\psi_\omega = 0). \tag{19}$$

Thus, putting $\mathbf{p} = -i\hbar\nabla$ in eq.(17) one obtains

$$\nabla^2\psi_\omega + n_\omega^2\frac{\omega^2}{c^2}\psi_\omega = 0. \tag{20}$$

The term $n_\omega^2 \omega^2$ mixes the kinetic and potential terms of the Schrödinger equation (16). From my point of view it is more natural to interpret the equation (17) in another way. That is, to put $E \equiv \hbar\omega$ and to rewrite the equation (17) as a connection between $E$ and $p$ in the form

$$E = \sqrt{c^2 p^2 + \frac{(\Omega_\omega - \Gamma_\omega)^2}{4}} - \frac{1}{2}(\Omega_\omega + \Gamma_\omega) . \tag{21}$$

It is apparent that $E$ is the energy of a massive relativistic particle in an external field. Thus the photon in dielectric gains the mass $m$ given by

$$m^2 c^4 \equiv \frac{(\Omega_\omega - \Gamma_\omega)^2}{4} . \tag{22}$$

The photon gains the mass, because of the interaction with the 'sea' of charges in dielectric. It reminds the Feynman's remark: „mass is interactions". The term

$$U = -\frac{1}{2}(\Omega_\omega + \Gamma_\omega) \tag{23}$$

is a 'classical' potential energy. It confirms the previous interpretation of the quantities $\Omega_\omega$ and $\Gamma_\omega$ as some potential energies. The equation (22) predicts that the mass of photon becomes zero not only in empty space (when $\Omega_\omega$ and $\Gamma_\omega$ vanish) but also when $\Omega_\omega = \Gamma_\omega$ (equivalently $\varepsilon_\omega = \mu_\omega$).

Thus the wave equation (17) takes on the form of the Klein-Gordon equation

$$(E - U)^2 \psi_\omega = (c^2 p^2 + m^2 c^4) \psi_\omega , \tag{24}$$

Certainly, the massive photon in dielectric has energy

$$E = \frac{mc^2}{\sqrt{1 - \frac{v^2}{c^2}}} + U . \tag{25}$$

Putting $E \equiv \hbar\omega$ and using (22), (23) one can calculate from (25) the velocity $v$ of photon in dielectric:

$$\frac{v^2}{c^2} = 1 - \frac{(\varepsilon_\omega - \mu_\omega)^2}{(\varepsilon_\omega + \mu_\omega)^2} \tag{26}$$

Note that $v$ is velocity of photon *as a particle*. It never exceeds $c$. The remarkable feature is that $v$ is equal to $c$ not only in empty space but also if $\varepsilon_\omega = \mu_\omega$ (as one should expect because then the photon mass is zero). Knowing $m$ and $v$ one can calculate the photon momentum

$$p = \frac{mv}{\sqrt{1 - \frac{v^2}{c^2}}} = n_\omega \frac{\hbar\omega}{c} . \tag{27}$$

The velocity $v$ is neither phase nor group velocity. The phase velocity $v_{ph}$ of photon is

$$v_{ph} \equiv \frac{\omega}{k} \equiv \frac{E}{p} = \frac{c}{n_\omega} , \qquad \text{(where } p \equiv \hbar k \text{)}. \tag{28}$$

And the group velocity $v_g$ is

$$v_g \equiv \frac{\partial \omega}{\partial k} \equiv \frac{\partial E}{\partial p} = \frac{c}{n_\omega + \omega \frac{\partial n}{\partial \omega}} . \tag{29}$$

In the *non-dispersive* case ($\partial n/\partial \omega = 0$) the group velocity is equal to the phase velocity. On the other hand the group velocity $v_g$ is equal to $v$ in the *independent* of $\omega$ case ($\partial \Omega/\partial \omega = 0$, $\partial \Gamma/\partial \omega = 0$). I think that one should consider the possibility that $v$ is really the true velocity of photon in dielectric. Certainly, the velocity of photon in dielectric medium is not the question of definition. The answer can give only an experiment.

Note that $\hbar \omega$ plays two *distinct* roles in the above description. It is the total energy and apart it is *a parameter* determining photon-medium interaction. It is the reason why the right hand side of the equation (21) depends on $\omega$.

At the end of this section I briefly comment the case of the wave packet

$$\int \psi_\omega d\omega = \int \varphi_\omega(\mathbf{r}) \exp(-i\omega t) d\omega . \tag{30}$$

For every Fourier component of the packet one may write Schrödinger equation (16) and thus Klein-Gordon equation (24). Because of different particle velocities $v$ the wave packet disperses. If the dispersion of velocities is $\Delta v$ the width of the packet is (in one dimension) increasing in time as $\Delta x = \Delta v \cdot t$. The packet describes *one* photon in a non-stationary state (the energy and the mass of the photon are not precisely determined) and $\Delta x$ is the region in which it is possible to detect it. Usually the beam of light contains many photons. It means that all the photons are in the same one-particle non-stationary state (30). Now, in the region $\Delta x$ you can detect many photons in different points at the same time. Probability is proportional to the rate of detection, and thus to the energy density in given point. You have a macroscopic quantum state.

## 4. THE VECTOR POTENTIAL OF THE MEDIUM

Developing analogy with the theory of charged particles it is interesting to construct and examine consequences of vector potential of the medium $A$ in the case of photon. Replacing $p \rightarrow P + A$, where $p$ is kinetical and $P$ canonical momentum, the Schrödinger equation (16) becomes

$$\hbar \omega \, \psi_\omega = c \, (P + A) \cdot S \begin{bmatrix} 1 & 0 \\ 0 & -1 \end{bmatrix} \psi_\omega - \Omega_\omega(\mathbf{r}) \, \gamma \, \psi_\omega - \Gamma_\omega(\mathbf{r}) \, \eta \, \psi_\omega \, . \tag{31}$$

To determine the situation, I suppose that the light source emitting photons of energies $\hbar \omega$ rests with respect to the observer. Then, as will be shown, the physical meaning of the vector potential of the medium is directly connected with the velocity $u$ of the medium. Note, there is no Doppler shift between the observer and the source and therefore the observed frequency is the same as the source frequency.

Expressing the Schrödinger equation (31) in the classical language one finds

$$\frac{\omega}{c} \boldsymbol{D} = i \nabla \times \boldsymbol{H}, \qquad \frac{\omega}{c} \boldsymbol{B} = -i \nabla \times \boldsymbol{E}; \qquad (32)$$

$$\boldsymbol{D} = \varepsilon_\omega \boldsymbol{E} + \boldsymbol{a} \times \boldsymbol{H}, \qquad \boldsymbol{B} = \mu_\omega \boldsymbol{H} - \boldsymbol{a} \times \boldsymbol{E},$$

where $\boldsymbol{a} = \dfrac{\boldsymbol{A}}{(\hbar \omega / c)}$ is a dimensionless vector potential of the medium.

This may be compared with the nonrelativistic approximation of the Minkowski relations [9-10] obliging for uniformly moving dielectric:

$$\boldsymbol{D} = \varepsilon_\omega \boldsymbol{E} + (\varepsilon_\omega \mu_\omega - 1)\, \vec{\beta} \times \boldsymbol{H}, \qquad \boldsymbol{B} = \mu_\omega \boldsymbol{H} - (\varepsilon_\omega \mu_\omega - 1)\, \vec{\beta} \times \boldsymbol{E}, \qquad (33)$$

where $\vec{\beta} = \boldsymbol{u}/c$. One finds immediately the connection between vector potential $\boldsymbol{a}$ and the velocity $\boldsymbol{u}$ of the medium

$$\boldsymbol{a} = (\varepsilon_\omega \mu_\omega - 1)\, \vec{\beta}. \qquad (34)$$

The result is in agreement with papers [7-8], where it has been obtained in another way.

If one wants to examine the purely relativistic velocities case, the form of the Schrödinger equation (31) must be changed. One reason is that in the nonrelativistic velocities case we assumed that the couplings with the medium are the same as in the case of the resting medium. It does not need be true. The second reason is that the moving medium in fact produces anisotropy of the whole system. This is not taken into account in the nonrelativistic velocities case. Therefore one should admit that the couplings for fields perpendicular and parallel with respect to the velocity of the medium $\vec{\beta}$ may be different. Thus one should consider the following Schrödinger equation

$$\hbar \omega\, \psi_\omega = c\,(\boldsymbol{P} + \boldsymbol{A}) \cdot \boldsymbol{S} \begin{bmatrix} 1 & 0 \\ 0 & -1 \end{bmatrix} \psi_\omega + \qquad (35)$$

$$- \Omega_{\perp\omega}(\mathbf{r}) \begin{bmatrix} E_\perp \\ E_\perp \end{bmatrix} - \Omega_{\|\omega}(\mathbf{r}) \begin{bmatrix} E_\| \\ E_\| \end{bmatrix} - \Gamma_{\perp\omega}(\mathbf{r}) \begin{bmatrix} iH_\perp \\ -iH_\perp \end{bmatrix} - \Gamma_{\|\omega}(\mathbf{r}) \begin{bmatrix} iH_\| \\ -iH_\| \end{bmatrix}.$$

Our task is to find the relativistic form of $\boldsymbol{A} = (\hbar\omega/c)\,\boldsymbol{a}$. I look for the solution in the form

$$\boldsymbol{a} = \alpha(\beta)\ \vec{\boldsymbol{\beta}}\ , \tag{36}$$

where $\alpha(\beta)$ is yet unknown function. Expressing the Schrödinger equation (35) in the classical language one obtains Maxwell equations with material relations in the form

$$\boldsymbol{D} = \varepsilon_{||\omega} \boldsymbol{E}_{||} + \varepsilon_{\perp\omega} \boldsymbol{E}_{\perp} + \alpha(\beta)\ \vec{\boldsymbol{\beta}} \times \boldsymbol{H}_{\perp}\ , \tag{37}$$

$$\boldsymbol{B} = \mu_{||\omega} \boldsymbol{H}_{||} + \mu_{\perp\omega} \boldsymbol{H}_{\perp} + \alpha(\beta)\ \vec{\boldsymbol{\beta}} \times \boldsymbol{E}_{\perp}\ .$$

($\varepsilon_{||}$ is defined as $1+\Omega_{||}/\hbar\omega$, etc.). The material relations should be compared with the relativistic form of the Minkowski relations [9-10]

$$\boldsymbol{D} = \varepsilon_{\omega} \boldsymbol{E}_{||} + \frac{1-\beta^2}{1-\varepsilon_{\omega}\mu_{\omega}\beta^2}\varepsilon_{\omega}\boldsymbol{E}_{\perp} + \frac{\varepsilon_{\omega}\mu_{\omega}-1}{1-\varepsilon_{\omega}\mu_{\omega}\beta^2}\vec{\boldsymbol{\beta}} \times \boldsymbol{H}_{\perp}\ , \tag{38}$$

$$\boldsymbol{B} = \mu_{\omega} \boldsymbol{H}_{||} + \frac{1-\beta^2}{1-\varepsilon_{\omega}\mu_{\omega}\beta^2}\mu_{\omega}\boldsymbol{H}_{\perp} - \frac{\varepsilon_{\omega}\mu_{\omega}-1}{1-\varepsilon_{\omega}\mu_{\omega}\beta^2}\vec{\boldsymbol{\beta}} \times \boldsymbol{E}_{\perp}\ ,$$

One immediately finds

$$\alpha(\beta) = \frac{\varepsilon_{\omega}\mu_{\omega}-1}{1-\varepsilon_{\omega}\mu_{\omega}\beta^2}\ , \tag{39}$$

and as well the other parameters as $\varepsilon_{||\omega}$, $\varepsilon_{\perp\omega}$, etc.

For a uniformly moving homogeneous medium the result (39) is exact. To a good approximation it can be useful as well in the case of nonuniformly moving inhomogeneous medium provided that the potentials $\Omega$, $\Gamma$ and the flow $\boldsymbol{u}$ vary only gradually, i.e., do not vary significantly over one optical wave length and one optical cycle.

For nonrelativistic velocities one can immediately write connection between energy and momentum simply by substitution in eq.(21) $\boldsymbol{p} \to \boldsymbol{P} + \boldsymbol{A}$

$$E = \sqrt{c^2(\boldsymbol{P}+\boldsymbol{A})^2 + \frac{(\Omega_{\omega}-\Gamma_{\omega})^2}{4}} - \frac{1}{2}(\Omega_{\omega}+\Gamma_{\omega})\ . \tag{40}$$

It is possible because (as one can easily check with the help of Maxwell equations (32)) the divergence condition $\nabla\cdot\boldsymbol{D} = 0$, $\nabla\cdot\boldsymbol{B} = 0$ is equivalent to the condition $\boldsymbol{p}\cdot\boldsymbol{\psi} = 0$ (similarly as it was in the case of the resting medium). The eq.(40) is exact for uniformly moving medium, and a good approximation in the case of nonuniformly moving inhomogeneous medium. Note that the mass of photon is the same as it was in resting medium.

If one wants to iterate the relativistic Schrödinger equation (35) it will be advantageous to write it in the more suitable form

$$\hbar\omega\,\psi_\omega = \left(H_f - \Omega_{\|\omega}R_\| \gamma - \Omega_{\perp\omega}R_\perp \gamma - \Gamma_{\|\omega}R_\| \eta - \Gamma_{\perp\omega}R_\perp \eta\right)\psi_\omega\,, \qquad (41)$$

where projection operators $R_\|$ and $R_\perp$ have been defined in the following way: $R_\| \mathbf{V} = \mathbf{V}_\|$ and $R_\perp \mathbf{V} = \mathbf{V}_\perp$ ($\mathbf{V}$ is a vector, $\|$ and $\perp$ with respect to the velocity of the medium). The $R_\|$ and $R_\perp$ commute with $\gamma$, $\eta$. The Hamiltonian $H_f$ has the same structure as in the case of resting medium but now $\mathbf{p} = \mathbf{P} + \mathbf{A}$. Another useful relations are $H_f \gamma + \gamma H_f = H_f$, $H_f \eta + \eta H_f = H_f$ and $H_f R_\| + R_\| H_f = H_f - H_f^\|$, $H_f R_\perp + R_\perp H_f = H_f + H_f^\|$. The Hamiltonian $H_f^\|$ differs from $H_f$ in such a way that $\mathbf{p}\cdot\mathbf{S}$ is replaced by $\mathbf{p}_\|\cdot\mathbf{S}$. In general the obtained result of iteration is rather complicated and I will not write it down. I only examine here the simplest but physically interesting case when the light in form of plane wave propagates in the same direction as the medium moves. In this case the result has exactly the same form as eq.(40) with the only difference that $\Omega_\omega$ is replaced by $\Omega_{\perp\omega}$ and $\Gamma_\omega$ by $\Gamma_{\perp\omega}$. In particular, the mass of photon $m_\beta$ in the relativistic case is given by

$$m_\beta c^2 \equiv \frac{|\Omega_{\perp\omega} - \Gamma_{\perp\omega}|}{2} = \frac{1}{2}\frac{1-\beta^2}{1-\varepsilon_\omega\mu_\omega\beta^2}|\varepsilon_\omega - \mu_\omega|\hbar\omega = \frac{1-\beta^2}{1-\varepsilon_\omega\mu_\omega\beta^2}mc^2\,, \qquad (42)$$

where $m = |\varepsilon_\omega - \mu_\omega|\hbar\omega/2c^2$ is the mass of photon in the rest medium. At first, it might seem queer that the mass has changed but it must be so, because the mass of photon in medium (never mind resting or moving) is always determined by the couplings ($\Omega$ and $\Gamma$) and these couplings in the moving medium have changed. The change of mass does not appear in the nonrelativistic velocities case because it is only the second order effect (with respect to $\beta$).

In the following I restrict considerations to the nonrelativistically moving media which, no doubt, is reasonable from practical point of view. It is interesting that in some limited sense it is possible to develop 'photodynamics' - describing behavior of photon, in close analogy to the electrodynamics - as the theory of charged particles. In particular one can determine a 'classical' force $F$ acting on photon

$$F = \dot{p} = \dot{P} + \dot{A} \ . \tag{43}$$

In the case of nonuniformly moving homogeneous medium from Hamiltonian (40) one finds

$$\dot{P} = -\frac{\partial E}{\partial \mathbf{r}} = -\frac{c^2 \mathbf{p} \cdot \frac{\partial A}{\partial \mathbf{r}}}{E - U} = -\mathbf{v} \cdot \frac{\partial A}{\partial \mathbf{r}} \ . \tag{44}$$

Here $E - U$ and $p$ are given by (25), (27) and $v$ it is the particle velocity of photon given by (26). The analogs of 'electric' $\widetilde{E}$ and 'magnetic' $\widetilde{H}$ fields can be defined

$$\widetilde{E} = \frac{\partial A}{\partial t} \ ; \quad \widetilde{H} = -\nabla \times A \ . \tag{45}$$

Therefore the force $F$ is nothing else but 'the optical Lorentz force'

$$F = \widetilde{E} + v \times \widetilde{H} \ . \tag{46}$$

In the more general case of nonuniformly moving inhomogeneous medium one obtains a bit more complicated result

$$\widetilde{E} = \frac{\partial A}{\partial t} - \frac{\partial U}{\partial \mathbf{r}} - \frac{\partial (mc^2)}{\partial \mathbf{r}} \sqrt{1 - \frac{v^2}{c^2}} = \frac{\partial A}{\partial t} + \frac{1}{2}\left\{\left(1 - \sqrt{1 - \frac{v^2}{c^2}}\right)\frac{\partial \Omega_\omega}{\partial \mathbf{r}} + \left(1 + \sqrt{1 - \frac{v^2}{c^2}}\right)\frac{\partial \Gamma_\omega}{\partial \mathbf{r}}\right\} \tag{47}$$

,

$$\widetilde{H} = -\nabla \times A \ .$$

Note that in inhomogeneous medium the couplings $\Omega_\omega$, $\Gamma_\omega$ depend on place, therefore the mass of photon is position-dependent. The fields $\widetilde{E}$ and $\widetilde{H}$ are gauge invariant. The change of the potentials $A$, $\Omega_\omega$, $\Gamma_\omega$ in the following way

$$\boldsymbol{A}' = \boldsymbol{A} + \nabla f \quad ; \quad \Omega'_\omega = \Omega_\omega - \frac{\partial f}{\partial t} \quad ; \quad \Gamma'_\omega = \Gamma_\omega - \frac{\partial f}{\partial t} \qquad (48)$$

does not change the fields (47). Here $f$ is whatever function of time and space.

## 5. SUMMARY

I find that the influence of the medium on photon can be described by some scalar and vector potentials. Scalar potentials are directly connected with permittivity and permeability of the medium, vector potential is connected with the velocity of the medium. The main novelty in the paper is that the notion of vector potential of the medium can be constructed also for relativistic velocities of the medium. Another new results are the formulas for the mass of photon in resting and moving dielectric and the velocity of photon as particle. The velocity is different from phase and group velocity. Quite interesting is the fact that the photon velocity is equal to $c$ not only in vacuum but also if $\varepsilon_\omega = \mu_\omega$. A consequence of describing the medium through scalar and vector potentials is existence of analogs of 'electric' and 'magnetic' fields, as well as the optical Lorentz force which describe the influence of the medium on the photon. The same as in the theory of charged particles the potentials are gauge fields and 'electric' and 'magnetic' fields are gauge invariant.

## ACKNOWLEDGEMENTS

I am very grateful to Prof. B. Jancewicz for a helpful correspondence and assistence in preparation of the paper for publication.

Author contact information:

Jarosław Zaleśny,

Institute of Physics, Technical University of Szczecin,

Al. Piastów 48, 70-310 Szczecin, Poland.

e-mail:

jarek@arcadia.tuniv.szczecin.pl